# A compact approach to higher-resolution resonant inelastic X-ray scattering detection using photoelectrons

Jan O. Schunck[1,2], Jens Buck[3,4], Robin Y. Engel[1,2,5], Simon R. Kruse[2], Simon Marotzke[1,4], Markus Scholz[1], Sanjoy K. Mahatha[3], Meng-Jie Huang[3], Henrik M. Rønnow[6], Georgi Dakovski[7], Moritz Hoesch[1], Matthias Kalläne[4,8], Kai Rossnagel[3,4,8], Martin Beye[1,2,5,*]

[1] Deutsches Elektronen-Synchrotron DESY, Notkestr. 85, 22607 Hamburg, Germany

[2] Physics Department, Universität Hamburg, Luruper Chaussee 149, 22761 Hamburg, Germany

[3] Ruprecht Haensel Laboratory, Deutsches Elektronen-Synchrotron DESY, Notkestr. 85, 22607 Hamburg, Germany

[4] Institut für Experimentelle und Angewandte Physik, Christian-Albrechts-Universität zu Kiel, Olshausenstr. 40, 24098 Kiel, Germany

[5] Department of Physics, AlbaNova University Center, Stockholm University, SE-10691 Stockholm, Sweden

[6] Laboratory for Quantum Magnetism, Ecole Polytechnique Fédérale de Lausanne (EPFL), CH-1015 Lausanne, Switzerland

[7] SLAC National Accelerator Laboratory, 2575 Sand Hill Road, Menlo Park, CA 94025, USA

[8] Ruprecht Haensel Laboratory, Christian-Albrechts-Universität zu Kiel, Olshausenstr. 40, 24098 Kiel, Germany

[*] Corresponding author: martin.beye@fysik.su.se

The detection of inelastically scattered soft X-rays with high energy resolution usually requires large grating spectrometers. Recently, photoelectron spectrometry for analysis of X-rays (PAX) has been rediscovered for modern spectroscopy experiments at synchrotron light sources. By converting scattered photons to electrons and using an electron energy analyser, the energy resolution for resonant inelastic X-ray scattering (RIXS) becomes decoupled from the X-ray spot size and instrument length. In this work, we develop PAX towards high energy resolution using a modern photoemission spectroscopy setup studying $Ba_2Cu_3O_4Cl_2$ at the Cu $L_3$-edge. We measure a momentum transfer range of 24% of the first Brillouin zone simultaneously. Our results hint at the observation of a magnon excitation below 100 meV energy transfer and show intensity variations related to the dispersion of *dd*-excitations. With dedicated setups, PAX can become an alternative to the best and largest RIXS instruments, while at the same time opening new opportunities to acquire RIXS at a range of momentum transfers simultaneously and combine it with angle-resolved photoemission spectroscopy in a single instrument.

# 1 Introduction

Resonant inelastic X-ray scattering (RIXS) has become a prime spectroscopic tool for the study of low-energy elementary excitations, including phonons and magnons [1,2]. Through the use of X-rays, core-hole resonances can be exploited to selectively probe the active site in complex materials [3]. The sufficiently large momentum of X-rays allows mapping the dispersion of collective excitations [4–8]. By exploiting the photon polarization, symmetry properties of excitations can be analysed [1,9–13].

However, achieving competitive energy resolution to resolve these phenomena at soft X-ray resonances remains challenging and requires special developments in monochromator beamlines and spectrometer instruments with state-of-the-art gratings, stable mechanics and advanced detectors [14–21]. Further resolution improvements are reaching technological limits in many components simultaneously. Current record resolution is achieved with spectrometers so large that they extend into annex buildings of synchrotron experimental halls [16,17,20].

In this work, we realize a different approach to high energy resolution RIXS termed photoelectron spectrometry for the analysis of X-rays (PAX). Competitive energy resolution with good signal levels is achieved in a compact instrument with a clear path for improvements beyond current records. Our setup also offers the advantage of simultaneous detection of a range of momentum transfers. Furthermore, the instrument can be easily adapted to perform angle-resolved photoemission spectroscopy (ARPES) as another very powerful X-ray spectroscopy technique, e.g., for quantum materials research [22–25].

The idea behind PAX is to convert X-ray photons scattered from the sample into photoelectrons, which are subsequently detected using a photoelectron spectrometer. This technique was pioneered more than 50 years ago [26–28] and has recently been rediscovered for the measurement of RIXS [29,30]. Scattered X-ray photons are converted into photoelectrons in a specially selected material which ideally shows sharp and intense photoemission lines. According to Einstein's formula $E_{\text{kin}} = h\nu - E_b - \phi$, the resulting electron kinetic energy $E_{\text{kin}}$ is a direct measure of the X-ray energy $h\nu$ as well as the electron binding energy $E_b$ in the converter for a material with a given work function $\phi$. The measured electron spectrum (the PAX spectrum) is thus a convolution of the RIXS emission spectrum from the sample and the photoemission spectrum of the converter. Rigorous characterization of the converter spectrum under identical measurement conditions makes it possible to recover the original RIXS spectrum using deconvolution algorithms.

Using experimental chambers designed for X-ray photoelectron spectroscopy (XPS), previous work [29,30] has identified advantages of PAX over grating spectrometers: PAX is cost-effective, compatible with photoelectron spectroscopy, and the energy resolution is decoupled from the X-ray spot size, making PAX suitable for experiments at high-intensity free-electron lasers, where sample damage at small spot sizes is a severe limitation. In this work, we explore the capability of this technique to achieve high energy resolution for RIXS experiments and to acquire a significant momentum transfer range simultaneously.

# 2 Concept

The converter material comprises the core component in our approach and therefore needs to be chosen with care: When measured under exactly the same conditions as the PAX spectrum, the converter spectrum contains all resolution-determining terms, and the true RIXS spectrum could in principle be deconvolved to any energy resolution. However, due to noise in the acquired spectra, the sharpest features in the measured converter spectrum determine the effectively achievable energy resolution. The converter spectrum resolution is limited by the lifetime broadening of the observed photoemission lines or temperature broadening when using the Fermi edge for deconvolution, as well as resolution limits in the beamline and electron spectrometer. While it is challenging to identify materials with core-level emission linewidths below 100 meV and high photoemission cross-sections in the soft X-ray range [30,31], using the Fermi edge of metallic converters can provide energy resolutions limited only by the converter temperature, albeit at rather low signal levels. Remaining resolution limits are then given by the monochromaticity of the incident beam (i.e., the beamline, with records in the range of 10 meV at the Cu $L$-edge [16,17,20]) and the energy resolution of the electron analyser (often reaching below 10 meV energy resolution [23,32–36]).

Resonant excitation of a sample with soft X-rays ejects electrons via photoemission processes as well as Auger decays of core excited states. Inelastic scattering as measured by RIXS is a minority process since fluorescence decays of the core excitations are typically less than one percent in the soft X-ray range [37]. To detect PAX signals, the overwhelming number of electrons emitted directly from the sample need to be blocked, while the emitted photons need to be transmitted to a converter.

In our approach, we integrate the functionalities of blocking direct electrons and converting the RIXS signal into a single, layered thin-film structure consisting of a support material coated with the converter film (here, e.g., several nanometres of Ag and Au). The converter faces the electron analyser. The thickness of the support is chosen such that it provides mechanical stability, blocks the direct electrons and also maximizes X-ray transmission to the converter film. By placing this device within a few millimetres of the sample, the collected solid angle for RIXS photons can be maximized.

Photons emitted from the sample in different directions are converted to electrons at different positions on the converter. By using the transmission mode of the electron analyser, different scattering angles, i.e., different momentum transfers, can be resolved simultaneously. For example, in our test geometry, the converter has a diameter of 4 mm and is placed at a distance of 3 mm from the sample. With this setup, we acquire scattering angles between 122° and 160°.

Below, we present a study of the antiferromagnetic insulator $Ba_2Cu_3O_4Cl_2$ (Ba2342), which is structurally similar to cuprate high-temperature superconductors. The characteristic RIXS spectra of Ba2342 are well-known [38] and show intense and slightly dispersive $dd$-excitation peaks between 1.5 eV and 3 eV energy loss, as well as a dispersive magnon excitation up to approximately 300 meV.

## 3 Experimental Setup

High-resolution PAX measurements are performed at the soft X-ray beamline P04 of the synchrotron storage ring PETRA III at DESY using the permanently installed ASPHERE III endstation (see, e.g., [39,40]), designed for XPS and ARPES studies. The X-ray beam is monochromatized using a 400 l/mm grating and an exit slit width varying between 20 µm and 100 µm. At the entrance of the experimental chamber, the beam is refocused with Kirkpatrick-Baez mirrors.

The Ba2342 sample, mounted on a special sample holder close to the PAX converter, is kept at approximately 15 K to 20 K (see Fig. 1). Electrons are analysed in a VG Scienta DA 30 hemispherical analyser. The sample was cleaved in air before insertion into vacuum. The surface was oriented such that the Cu-O planes extended along the sample surface directions and the $a$-axis of the crystal was in the scattering plane. The incident photon energy was tuned to the absorption maximum of the Cu $L_3$ absorption edge at 931 eV and the polarization was circular.

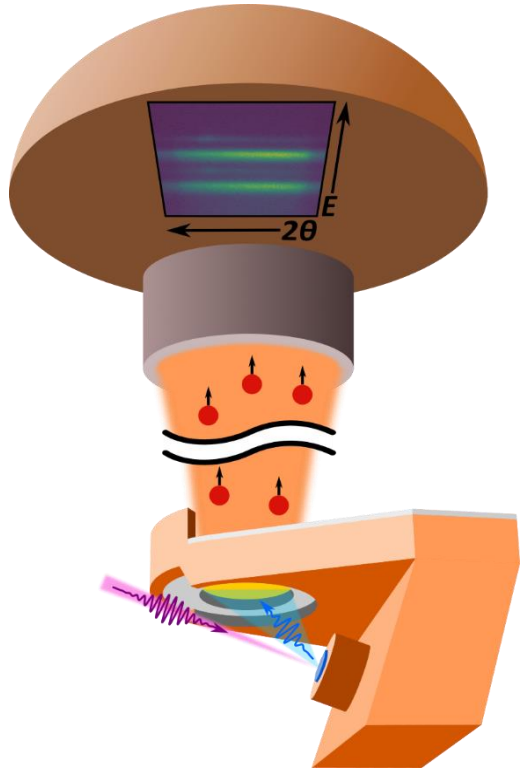

Figure 1: **Schematic of the PAX setup implemented into a hemispherical electron spectrometer.** Incident X-rays (purple) scatter (blue) from the sample and are subsequently converted into photoelectrons (red circles) using a transmission converter consisting of a thin metallic gold or silver layer (yellow). The photoelectron spectra (energy axis $E$) are measured using the transmission mode of a hemispherical electron spectrometer that radially disperses electrons according to their energy. Furthermore, electrons originating from different locations on the converter, corresponding to different scattering angles $2\theta$, can be imaged onto the two-dimensional detector, allowing to record of a range of momentum transfers from X-rays to the sample.

Measurements were performed with two different converter designs. For both converter designs, different beamline analyser settings were chosen as a compromise between the achievable resolution and signal level. The first configuration used a 200 nm polyimide support and a 10 nm gold film as converting material grown on a 3 nm titanium buffer layer. We used the Au $4f$ lines for conversion and employed the following resolution settings: the beamline exit slit was set to 50 µm, yielding a theoretical resolution of approximately 250 meV at the Cu $L$-edge, and the analyser slit was set to 300 µm, giving a theoretical electron energy resolution of 150 meV. For the second converter design, we used silver $3d$ lines for the conversion, which are naturally sharper compared to the gold $4f$ lines and could thus enable a more reliable deconvolution to high resolutions. On a 200 nm aluminium substrate, a 30 nm Ag converter layer was grown on top of a 20 nm organic buffer layer. The beamline exit slit was set to 20 µm (about 130 meV theoretical resolution) and the analyser entrance slit to 100 µm, giving 50 meV theoretical electron energy resolution. To efficiently acquire a sufficiently large energy window over extended times, we set the analyser pass energy to 200 eV and fixed the central analysed electron kinetic energy.

Panels (a) and (c) in Fig. 2 show the measured PAX spectra using the Au $4f$ and Ag $3d$ levels. The results obtained with the gold converter are shown in the Fig. 2 (a) and (b), and those obtained with the silver converter are shown in the Fig. 2 (c) and (d).

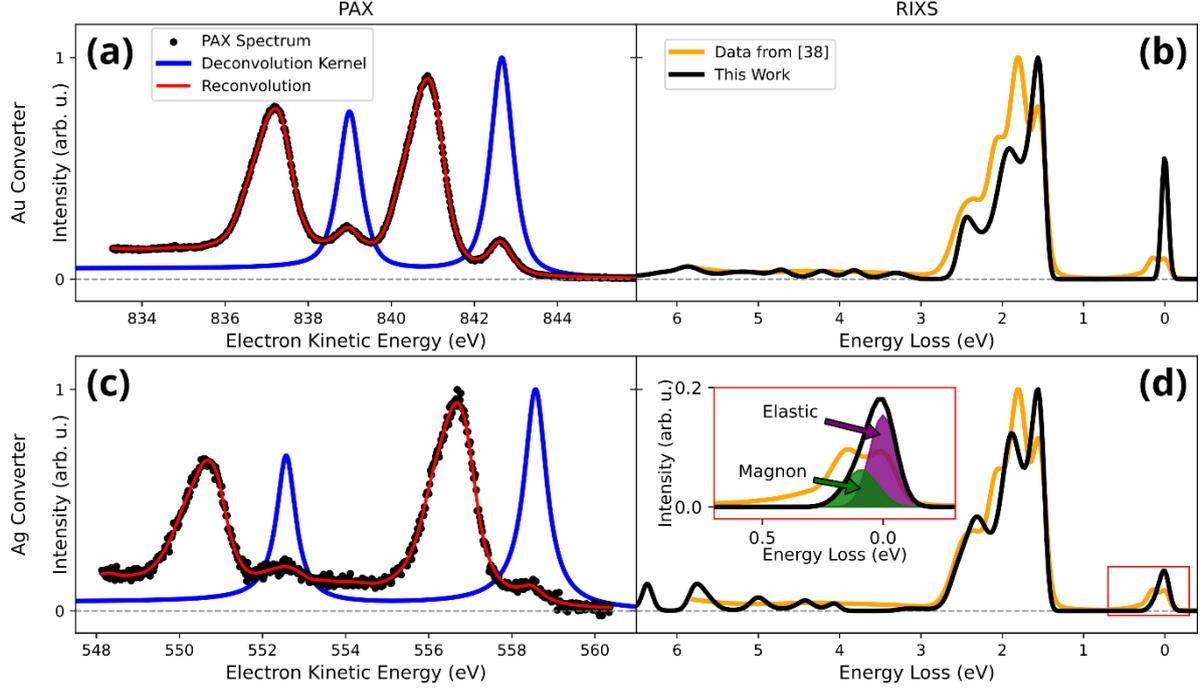

Figure 2: **PAX data and deconvolution results.** Results for measurements using a gold converter (panels a and b) and a silver converter (panels c and d). Moreover, panels (a) and (c) depict the measured PAX spectrum (black dots), converter photoemission used as deconvolution kernel (blue) and reconvolution (red) of this kernel with the spectrum resulting from the deconvolved RIXS spectra (black lines in panels (b) and (d)). Panels (b) and (d) furthermore show deconvolved Ba2342 RIXS spectra (black) compared to data taken at the ADRESS beamline [38] (orange). The inset in panel (d) depicts fits to two Gaussians representing the elastic line (purple) and the magnon peak (green). The RIXS spectra are momentum-integrated within the momentum ranges that are shown in Fig. 3, i.e. the data from this work (black) is integrated from $0.22\pi$ to $0.46\pi$ and the data from [38] (orange) is integrated from $0.18\pi$ to $0.53\pi$ (see also Supplementary Information).

To obtain the pure converter photoemission spectrum, which determines the kernel for the deconvolution of the PAX spectrum, our test setup allowed for two options: the most efficient is to place the converter directly in the beam. Besides the obstacle of necessarily attenuating the signal levels by many orders of magnitude without changing the spectral properties (due to much higher efficiencies of direct photoemission compared to PAX), this option is further challenged by an intrinsically different geometry from the PAX measurements, which affects the spectral properties by introducing systematic errors to the deconvolution. For PAX, the focus of the X-ray beam is placed on the sample, while the focus of the analyser should be on the converter. In our geometry, the two foci are 3 mm apart, posing a major challenge for the alignment of X-ray and electron optics, since our test setup was originally not designed for coinciding foci. As consequence, aberrations can affect the spectra. However, for the described direct photoemission measurement of the converter, the foci may coincide, changing the optical conditions and altering the spectral influence of the aberrations thus hampering successful deconvolution. We therefore employ a different strategy to measure the converter spectrum: Retaining the PAX geometry, we employ an elastic scatterer close to the sample, so that an unaltered converter photoemission spectrum can be measured without significant change in geometry. However, the signal level in this case is much lower than in a direct measurement.

For PAX with the Au converter, we measured the Au $4f_{5/2}$ and $4f_{7/2}$ core-level photoemission spectrum by illuminating an elastically scattering part of the sample holder next to the Ba2342 sample. For the PAX measurements, we moved the sample into the X-ray beam. In our test setup with the converter fixed to the sample holder, this thus moves both components and also slightly changes the converter position towards the electron analyser, with minor effects on the spectra. The blue line in panel (a) of Fig. 2 shows the fit to the converter Au 4*f* spectrum using two pseudo-Voigt peaks (see Supplementary Information). The measured PAX spectra (black dots) show two small and two large, asymmetric peaks. These doublets result from the convolution of the Ba2342 RIXS spectrum with the spin-orbit split converter spectra. The small peaks originate from elastic scattering and low-energy transfer excitations of the Ba2342 excitation spectrum, while the large peaks originate from the *dd* excitations [38]. The orange line in panel (b) is the Ba2342 RIXS spectrum as previously published [38]. The black line in panel (b) shows our RIXS spectrum as a deconvolution of the PAX spectrum. The red line in (a) is a reconstructed PAX spectrum for consistency check, obtained by again convolving the extracted RIXS spectrum with the converter spectrum. The absence of deviations within the noise level between the reconstructed PAX spectrum and the measured black dots demonstrates that the remaining differences between the extracted RIXS spectra and the true spectra are beyond the systematic and statistical limits of the current data set.

Corresponding spectra for the silver converter are shown in panels (c) and (d) of Fig. 2. A different Ba2342 crystal was used for these measurements. The observed photon-electron conversion originated from the silver $3d_{3/2}$ and $3d_{5/2}$ levels. The measurement time for all spectra was several hours.

To extract RIXS spectra from the PAX measurements, we use iterative deconvolution algorithms that have been benchmarked for this purpose in [27,29,30]. In order to suppress the artificial amplification of high frequencies in the process (which would, e.g., exceed the energy resolution limits), the estimated RIXS spectrum is regularized by convolution with a Gaussian after each iteration (see the Supplementary Information for details) [30]. Due to the limited measured energy window, it was challenging to properly subtract backgrounds from secondary electron scattering, optical aberrations, as well as inhomogeneities of the analyser transfer function and the detector. Their proper treatment is important for a high-quality deconvolution. Our procedure to account for these effects is detailed in the Supplementary Information.

We emphasize, however, that we use a deconvolution algorithm that does not require any prior knowledge of the spectrum. Each intensity point in the extracted spectrum is treated as a free parameter and the deviation between the measured PAX spectrum and the extracted RIXS spectrum after convolution with the kernel is minimized. The only external inputs are the regularization strength, which limits the spectral width of the features to experimentally reasonable values and thus gains physical meaning, and the number of iterations.

## 4 Results and Discussion

Generally, the main features of the reference spectrum (orange lines in Figs. 2 (b) and 2 (d)) are well reproduced by the extracted RIXS spectra. As is common in RIXS measurements, the intensity of the elastic line varies between samples and measurement positions and is found to be more intense from our sample next to the Au converter. The elastic peak in the measurement with the Ag converter is found to be asymmetric, which is connected to the magnon excitation in Ba2342 at energies up to approximately 300 meV. The extracted feature can be fitted with two Gaussian peaks shifted by 91 meV with widths of 130 meV (FWHM) for the elastic peak and 205 meV for the energy loss feature, as shown in the inset of Fig. 2 (d). The absence of this asymmetry in the Au converter data is related to the generally lower resolution and the stronger elastic line in this data set. Due to the lower resolution and higher elastic line strength, this feature is not resolved in the Au spectra. In general, spectra after deconvolution can display unphysically sharp features, sharper than supported by the quality of the measured data.

The general shape of the *dd*-excitation signatures between 1.5 eV and 3 eV energy loss is reproduced by both estimated RIXS spectra. However, the deconvolution results in three peaks of decreasing intensity with energy loss, while four peaks were previously reported [38]. The peak at the lowest energy loss in this report is less intense, but the other peaks also decrease in intensity with energy loss as in our data.

In the region of charge transfer excitations with energy losses greater than 3 eV, we find spectral weight from PAX as well as the reference spectrum. While the reference spectrum displays a smooth broad distribution, the PAX deconvolution concentrates the spectral weight into several sharper features. However, we note that our limited detection window in this experiment does not fully cover this energy loss region with both core levels from the converter. This, together with the low charge transfer intensity, makes this region susceptible to generating artefacts during the deconvolution.

In this pioneering experiment, we aimed to achieve a high energy resolution in the extracted RIXS spectrum, beyond the total resolution of the instrument. We therefore kept the regularization strength slightly lower than suggested in previous studies [30] in order to resolve sharp features. This led to the successful observation of the magnon excitation also with this method but at the expense of spurious sharpness in the charge transfer spectral region and uncertainties in the number and intensity ratio of the main *dd* excitations. Other reasons for this difference are residual uncorrected background values and slopes, as well as the limited energy detection window in the fixed analyser setup used.

Nevertheless, some of the differences between our approach and the reference may not be artefacts of our procedure, but may indeed be real due to a different sample growth and treatment and a different measurement geometry as compared to the reference data. Unfortunately, the experimental constraints of our setup do not yet allow for a more rigorous conclusion, but these ambiguities can be resolved with a specifically designed instrument that addresses limitations in resolution, avoids aberrations due to different foci, acquires a larger energy range, and has a more homogeneous transfer and detection function in the electron analyser.

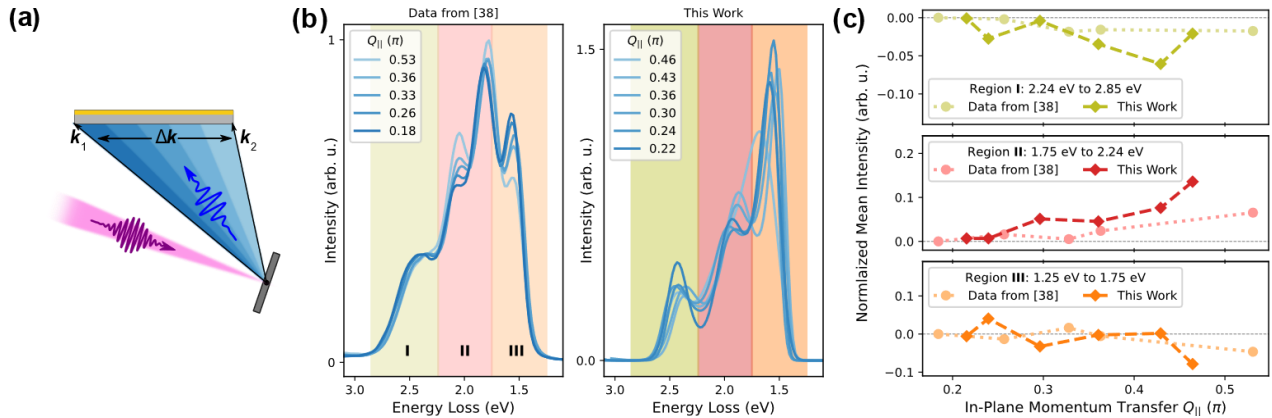

Figure 3: **Momentum-resolved PAX.** **(a)** Incident X-rays (purple) emit a fan of scattered X-rays (blue) with different wavevectors (e.g., $k_1$ and $k_2$) impinging on different positions on the converter, thereby translating momentum transferred from the sample to scattered X-rays into spatial coordinate on the converter. **(b)** RIXS data from the ADRESS beamline [38] (left) and RIXS deconvolved from PAX measurements (right) for a range of in-plane momentum transfers $Q_\parallel$. The energy range of the *dd* excitations is divided into three regions I, II and III, highlighted by the green, red and orange shading, respectively. **(c)** The mean intensities in the coloured regions of the spectra from the reference and this work (shown in panel (b)) are normalized and compared. See Supplementary Information for a more detailed description of the analysis of the momentum-resolved spectra.

A major advantage of PAX over RIXS measured with a grating spectrometer is the ability to acquire a large range of scattering angles simultaneously (see Fig. 3). In our setup, a scattering angle range from 122° to 160° is covered, corresponding to an in-plane momentum transfer $Q_\parallel$ range from $0.22\pi$ to $0.46\pi$, or about 24% of the first Brillouin zone. To assess this capability, we compare the evolution of spectral weight within the energy loss region of *dd*-excitations between reference RIXS spectra [38] (Fig. 3(b), left) and our extracted spectra from PAX using a gold converter by dividing the 2D detector image equally into six parts along the non-dispersive direction (Fig. 3(b) right). All spectra are normalized to the integrated intensity within the energy range shown.

The energy loss region is further divided into three characteristic regions between 1.5 eV and 2.5 eV energy loss, indexed I to III. The integral intensity in each region is compared as a function of the momentum transfer $Q_\parallel$ after subtracting the value at the lowest $Q_\parallel$ (Fig. 3(c)). While the intensities in region I display a decreasing intensity with $Q_\parallel$, we observe an increasing intensity with $Q_\parallel$ in region II, and intensity roughly independent of $Q_\parallel$ in region III. These trends are qualitatively observed both in the reference spectra measured with a grating spectrometer in sequential mode (circles) and in our parallel measurement with PAX (diamonds). We note, however, that our PAX measurement was affected by aberrations of the analyser imaging, which can be improved in a dedicated setup.

## 5 Conclusion

In summary, we demonstrate on the one hand the capabilities of PAX to achieve high energy resolution for RIXS at competitive signal levels using a repurposed test setup. Furthermore, the simultaneous momentum-resolved RIXS acquisition of relevant ranges of momentum transfers presents a step beyond grating-based scattering experiments. With this data set, we provide a compelling case for investing in a dedicated setup that will overcome current limitations and provide an instrument with energy resolution and efficiency competitive with large grating spectrometers at a much smaller size and cost, which can naturally also be used for modern electron spectroscopy, such as ARPES.

## 6 Acknowledgements

We thank Marco Grioni for providing us with the raw data for the Ba2342 RIXS reference spectra (ref [38]). Furthermore, we thank Pablo J. Bereciartua for the alignment measurements of the sample crystals using Laue diffraction. We acknowledge DESY (Hamburg, Germany), a member of the Helmholtz Association HGF, for the provision of experimental facilities. Parts of this research were carried out at PETRA III. Funding for the photoemission spectroscopy instrument at beamline P04 (Contracts 05KS7FK2, 05K10FK1, 05K12FK1, 05K13FK1, 05K19FK4 with Kiel University; 05KS7WW1, 05K10WW2 and 05K19WW2 with Würzburg University) by the Federal Ministry of Education and Research (BMBF) is gratefully acknowledged. Beamtime was allocated for proposal(s) I-20191064. J.O.S., R.Y.E. and M.B. were supported by the Helmholtz Association through grant VH-NG-1105.

## 7 References

[1] L. J. P. Ament, M. van Veenendaal, T. P. Devereaux, J. P. Hill, and J. van den Brink, *Resonant Inelastic X-Ray Scattering Studies of Elementary Excitations*, Rev. Mod. Phys. **83**, 705 (2011).

[2] A. Kotani and S. Shin, *Resonant Inelastic X-Ray Scattering Spectra for Electrons in Solids*, Rev. Mod. Phys. **73**, 203 (2001).

[3] F. J. Himpsel, *Photon-in Photon-out Soft X-ray Spectroscopy for Materials Science*, Phys. Status Solidi **248**, 292 (2011).

[4] M. P. M. Dean et al., *Spin Excitations in a Single $La_2CuO_4$ Layer*, Nat. Mater. **11**, 850 (2012).

[5] Y. Y. Peng et al., *Influence of Apical Oxygen on the Extent of In-Plane Exchange Interaction in Cuprate Superconductors*, Nat. Phys. **13**, 1201 (2017).

[6] L. Chaix et al., *Dispersive Charge Density Wave Excitations in $Bi_2Sr_2CaCu_2O_{8+\delta}$*, Nat. Phys. **13**, 952 (2017).

[7] G. Ghiringhelli et al., *Crystal Field and Low Energy Excitations Measured by High Resolution RIXS at the $L_3$ Edge of Cu, Ni and Mn*, Eur. Phys. J. Spec. Top. **169**, 199 (2009).

[8] L. Braicovich et al., *Magnetic Excitations and Phase Separation in the Underdoped $La_{2-x}Sr_xCuO_4$ Superconductor Measured by Resonant Inelastic X-Ray Scattering*, Phys. Rev. Lett. **104**, 3 (2010).

[9] A. Kotani, *Resonant Inelastic X-Ray Scattering and Its Magnetic Circular Dichroism*, J. Phys. Chem. Solids **66**, 2150 (2005).

[10] L. Braicovich, G. Van Der Laan, G. Ghiringhelli, A. Tagliaferri, M. A. Van Veenendaal, N. B. Brookes, M. M. Chervinskii, C. Dallera, B. De Michelis, and H. A. Dürr, *Magnetic Circular Dichroism in Resonant Raman Scattering in the Perpendicular Geometry at the L Edge of 3d Transition Metal Systems*, Phys. Rev. Lett. **82**, 1566 (1999).

[11] A. Nag et al., *Many-Body Physics of Single and Double Spin-Flip Excitations in NiO*, Phys. Rev. Lett. **124**, 067202 (2020).

[12] H. Elnaggar, A. Nag, M. W. Haverkort, M. Garcia-Fernandez, A. Walters, R.-P. Wang, K.-J. Zhou, and F. de Groot, *Magnetic Excitations beyond the Single- and Double-Magnons*, Nat. Commun. **14**, 2749 (2023).

[13] J. Li et al., *Single- and Multimagnon Dynamics in Antiferromagnetic $AFe_2O_3$*, Phys. Rev. X **13**, 011012 (2023).

[14] F. Gel'mukhanov and H. Ågren, *Resonant X-Ray Raman Scattering*, Phys. Rep. **312**, 87 (1999).

[15] C. Schulz, K. Lieutenant, J. Xiao, T. Hofmann, D. Wong, and K. Habicht, *Characterization of the Soft X-Ray Spectrometer PEAXIS at BESSY II*, J. Synchrotron Radiat. **27**, 238 (2020).

[16] N. B. Brookes et al., *The Beamline ID32 at the ESRF for Soft X-Ray High Energy Resolution Resonant Inelastic X-Ray Scattering and Polarisation Dependent X-Ray Absorption Spectroscopy*, Nucl. Instruments Methods Phys. Res. Sect. A Accel. Spectrometers, Detect. Assoc. Equip. **903**, 175 (2018).

[17] K.-J. Zhou et al., *I21: An Advanced High-Resolution Resonant Inelastic X-Ray Scattering Beamline at Diamond Light Source*, J. Synchrotron Radiat. **29**, 563 (2022).

[18] V. N. Strocov et al., *High-Resolution Soft X-Ray Beamline ADRESS at the Swiss Light Source for Resonant Inelastic X-Ray Scattering and Angle-Resolved Photoelectron Spectroscopies*, J. Synchrotron Radiat. **17**, 631 (2010).

[19] G. Ghiringhelli et al., *SAXES, a High Resolution Spectrometer for Resonant x-Ray Emission in the 400–1600eV Energy Range*, Rev. Sci. Instrum. **77**, (2006).

[20] J. Dvorak, I. Jarrige, V. Bisogni, S. Coburn, and W. Leonhardt, *Towards 10 meV Resolution: The Design of an Ultrahigh Resolution Soft X-Ray RIXS Spectrometer*, Rev. Sci. Instrum. **87**, (2016).

[21] A. Singh et al., *Development of the Soft X-Ray AGM-AGS RIXS Beamline at the Taiwan Photon Source*, J. Synchrotron Radiat. **28**, 977 (2021).

[22] Z.-X. Shen and D. S. Dessau, *Electronic Structure and Photoemission Studies of Late Transition-Metal Oxides — Mott Insulators and High-Temperature Superconductors*, Phys. Rep. **253**, 1 (1995).

[23] A. Damascelli, Z. Hussain, and Z.-X. Shen, *Angle-Resolved Photoemission Studies of the Cuprate Superconductors*, Rev. Mod. Phys. **75**, 473 (2003).

[24] P. D. C. King, S. Picozzi, R. G. Egdell, and G. Panaccione, *Angle, Spin, and Depth Resolved*

*Photoelectron Spectroscopy on Quantum Materials*, Chem. Rev. **121**, 2816 (2021).

[25] J. A. Sobota, Y. He, and Z.-X. Shen, *Angle-Resolved Photoemission Studies of Quantum Materials*, Rev. Mod. Phys. **93**, 025006 (2021).

[26] M. O. Krause, *Determination of L Binding Energies of Krypton by the Photoelectron Method*, Phys. Rev. **140**, A1845 (1965).

[27] M. F. Ebel, *On the Use of an Electron Spectrometer as Detector for Soft X-Ray Spectra*, X-Ray Spectrom. **4**, 43 (1975).

[28] O. Benka, *A Cylindrical Mirror Photoelectron Spectrometer with Position-Sensitive Detector Used for X-Ray Analysis*, Nucl. Instruments Methods Phys. Res. **203**, 547 (1982).

[29] G. L. Dakovski, M.-F. Lin, D. S. Damiani, W. F. Schlotter, J. J. Turner, D. Nordlund, and H. Ogasawara, *A Novel Method for Resonant Inelastic Soft X-Ray Scattering via Photoelectron Spectroscopy Detection*, J. Synchrotron Radiat. **24**, 1180 (2017).

[30] D. J. Higley, H. Ogasawara, S. Zohar, and G. L. Dakovski, *Using Photoelectron Spectroscopy to Measure Resonant Inelastic X-Ray Scattering: A Computational Investigation*, J. Synchrotron Radiat. **29**, 202 (2022).

[31] A. Thompson et al., *X-Ray Data Booklet*, J. Synchrotron Radiat. **8**, 1125 (2001).

[32] V. N. Strocov, X. Wang, M. Shi, M. Kobayashi, J. Krempasky, C. Hess, T. Schmitt, and L. Patthey, *Soft-X-Ray ARPES Facility at the ADRESS Beamline of the SLS: Concepts, Technical Realisation and Scientific Applications*, J. Synchrotron Radiat. **21**, 32 (2014).

[33] A. V Fedorov, T. Valla, P. D. Johnson, Q. Li, G. D. Gu, and N. Koshizuka, *Temperature Dependent Photoemission Studies of Optimally Doped $Bi_2Sr_2CaCu_2O_8$*, Phys. Rev. Lett. **82**, 2179 (1999).

[34] T. Yokoya, T. Kiss, T. Watanabe, S. Shin, M. Nohara, H. Takagi, and T. Oguchi, *Ultrahigh-Resolution Photoemission Spectroscopy of Ni Borocarbides: Direct Observation of the Superconducting Gap and a Change in Gap Anisotropy by Impurity*, Phys. Rev. Lett. **85**, 4952 (2000).

[35] N. P. Armitage et al., *Superconducting Gap Anisotropy in $Nd_{1.85}Ce_{0.15}CuO_4$: Results from Photoemission*, Phys. Rev. Lett. **86**, 1126 (2001).

[36] M. Hoesch et al., *A Facility for the Analysis of the Electronic Structures of Solids and Their Surfaces by Synchrotron Radiation Photoelectron Spectroscopy*, Rev. Sci. Instrum. **88**, (2017).

[37] M. O. Krause, *Atomic Radiative and Radiationless Yields for K and L Shells*, J. Phys. Chem. Ref. Data **8**, 307 (1979).

[38] S. Fatale et al., *Electronic and Magnetic Excitations in the Half-Stuffed Cu-O Planes of $Ba_2Cu_3O_4Cl_2$ Measured by Resonant Inelastic x-Ray Scattering*, Phys. Rev. B **96**, 115149 (2017).

[39] M. Ünzelmann et al., *Momentum-Space Signatures of Berry Flux Monopoles in the Weyl Semimetal TaAs*, Nat. Commun. **12**, 3650 (2021).

[40] Y. Kato et al., *Electronic Structure of $ACu_3Co_4O_{12}$ (A = Y, La, Bi): Synthesis, Characterization, Core-Level Spectroscopies, High-Pressure Application, and Ab Initio Calculation*, Phys. Rev. Mater. **7**, 073401 (2023).